# A Model of Zebra Patterns in Solar Radio Emission

G. P. Chernov[a,] *, V. V. Fomichev[a], and R. A. Sych[b]

[a]*Pushkov Institute of Terrestrial Magnetism, Ionosphere, and Radiowave Propagation, Russian Academy of Sciences (IZMIRAN), Troitsk, Moscow, 108840 Russia*

[b]*Institute of Solar and Terrestrial Physics, Siberian Division, Russian Academy of Sciences, Irkutsk, 664033 Russia *e-mail: gchernov@izmiran.ru*

**Abstract**—We analyze complex zebra patterns and fiber bursts during type-IV solar radio bursts on August 1, 2010. It was shown that all of the main details of sporadic zebra patterns can be explained within the model of zebra patterns and fiber bursts during the interaction of plasma waves with whistlers. In addition, it was shown that the major variations in the stripes of the zebra patterns are caused by the scattering mechanism of fast particles on whistlers, which leads to the transition of whistler instability from the normal Doppler effect to an anomalous one.

1. INTRODUCTION

Though there are ten suggested mechanisms, zebra patterns (ZPs) in the form of periodical emission and absorption bands in the dynamic spectrum of radio emission remain the most intriguing fine structure of type-IV solar continuous radio bursts (Chernov, 2011). This is primarily related to the wide variety of stripes in each new phenomenon, when it becomes impossible to explain all fine details based on a single mechanism. The most widespread explanation is the emission at different levels of double plasma resonance (DPR), when the upper hybrid frequency ($\omega_{UH}$) becomes equal to the whole number of electronic cyclotron harmonics $s\omega_{Be}$: $\omega_{UH} = (\omega^2_{Pe} + \omega^2_{Be})^{1/2} = s\omega_{Be}$. All aspects of the DPR mechanism were described in a recent review (Zheleznyakov et al., 2016). In this model, all variations in zebra stripes are usually associated with variations in the magnetic field in the source, in particular, with the propagating magnetic acoustic waves. In special complex cases, various combinations of the density and magnetic field gradients are assumed or a simultaneous presence of three different functions of the distribution of fast particles exists in the source. For example, the realization of this mechanism was justified (Chernov et al., 1998) for an individual, slowly drifting fiber. The DPR mechanism was even suggested for the interpretation of stripes similar to ZPs in the microwave radio emission of a pulsar in the Crab Nebula, despite the significant inconsistence between the parameters of the zebra stripes and numerous problems with the plasma parameters as compared with the solar parameters (Zheleznyakov et al., 2016).

New calculations of the increments of upper hybrid waves under DPR conditions (Kuznetsov and Tsap, 2007) showed that the maximum of increments providing realistic ZP bands can be obtained if the spectrum of fast particles has a power shape with a large exponent (~8).

A number of authors continue to improve the mechanism based on DPR. It was shown (Karlický and Yasnov, 2015) that ZPs can be excited within the DPR in the transition layer only in a narrow interval of altitudes with increased density and high gradient (a model described by Selhorst et al. (2008) and Yasnov et al. (2016)), but without any explanation of the formation of this narrow density peak. The zebra frequency range is limited on the high-frequency side by bremsstrahlung, in the low frequency band by the absorption on the cyclotron harmonics. It was also shown that it is not possible to use the barometric density formula in the zebra sources.

New calculations of increments of upper hybrid waves under DPR conditions with a ring distribution of fast electrons were performed in a recent publication (Benáček et al., 2017) in which the authors used relativistic corrections for different temperatures of the background plasma and fast particles. It was shown that a clear increment maximum exists only if the electron velocities are 0.1 $c$ with a narrow dispersion. At a velocity of 0.2 $c$, the increment sharply decreases and the maxima already merge in the continuum for a few cyclotron harmonics $s$. In this case, it is not taken into account that the zebra stripes yet show a superfine structure of millisecond duration. Under such conditions it becomes clear that realization of the DPR mechanism faces many problems, including the provision of a large number of zebra stripes and the sporadic character of their appearance.

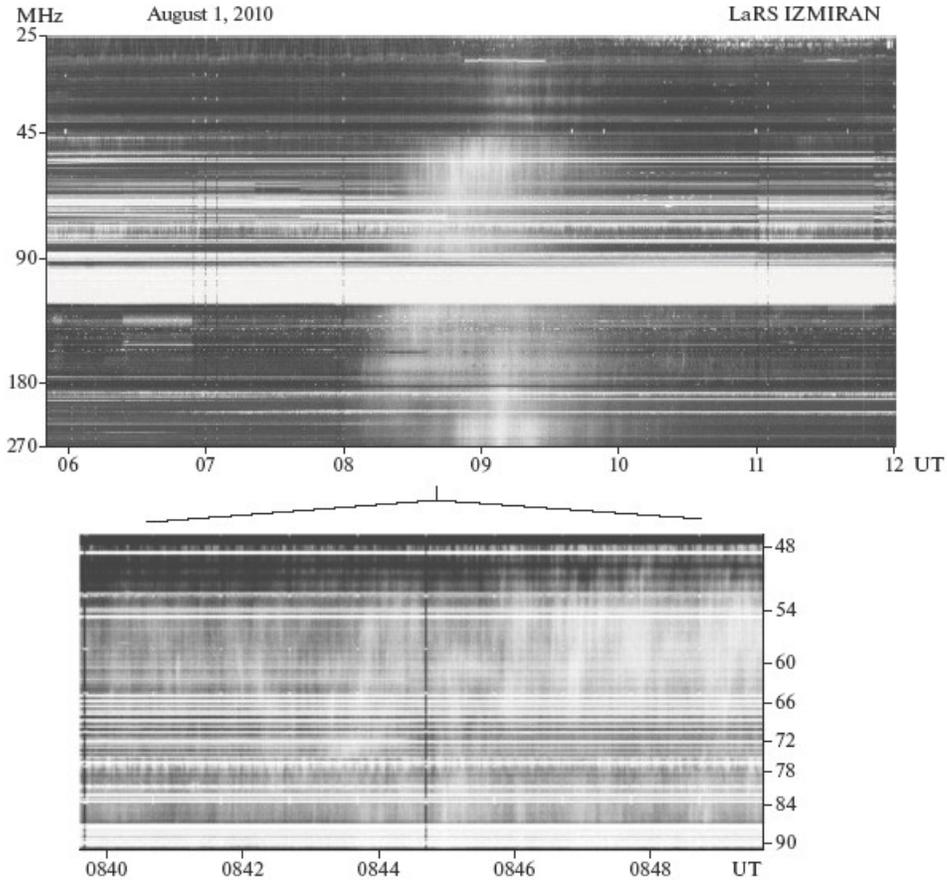

Fig. 1. Radio burst on August 1, 2010, based on the data from the IZMIRAN spectrograph in the range 25−270 MHz (upper panel). Increased spectrum fragment in the range 45–90 MHz is shown in the bottom panel. The burst has the form of slowly drifting continuum. No clearly manifested type-II burst stripes are seen, which coincides with the Hiraiso spectrum.

The phenomenon of August 1, 2010, considered here was rich in the fine structure of radio emission in the entire range from 25 to 3500 MHz: in addition to the fast pulsations in the meter range, a wide variety of ZPs and fiber bursts were observed in the decimeter and microwave ranges.

Several spectra of this phenomenon measured at the Ondřejov observatory in the decimeter range were presented (Karlický, 2014); the stripes of the ZPs reveal a superfine structure with a periodicity of ~30 ms (the spectrograph resolution is 10 ms). The author continues and analyzes this work using the DPR mechanism in turbulent plasma (Karlický et al., 2001). The variations in the HF boundary of the zebra stripes are calculated. In fact, they continue the work (Karlický et al., 2001) to interpret the lace bursts in turbulent plasma with resonance intensification of the radiation of two oscillators at frequencies $\omega_{UH}$ and $s\omega_{Be}$ with a positive feedback. The magnetic field is considered constant, while the density changes with a high periodicity (up to 30 Hz) due to the propagation of a wave with the sound velocity. The alternation of individual spikes is not considered a superfine structure. Using a series of burst images from a space instrument SDO/AIA at a wavelength of 171 Å in the extreme ultra violet emission, Yasnov et al. (2016) determine the plasma parameters of the radio source for the next moment of a ZP in the event on August 1, 2010, within the framework of the DPR.

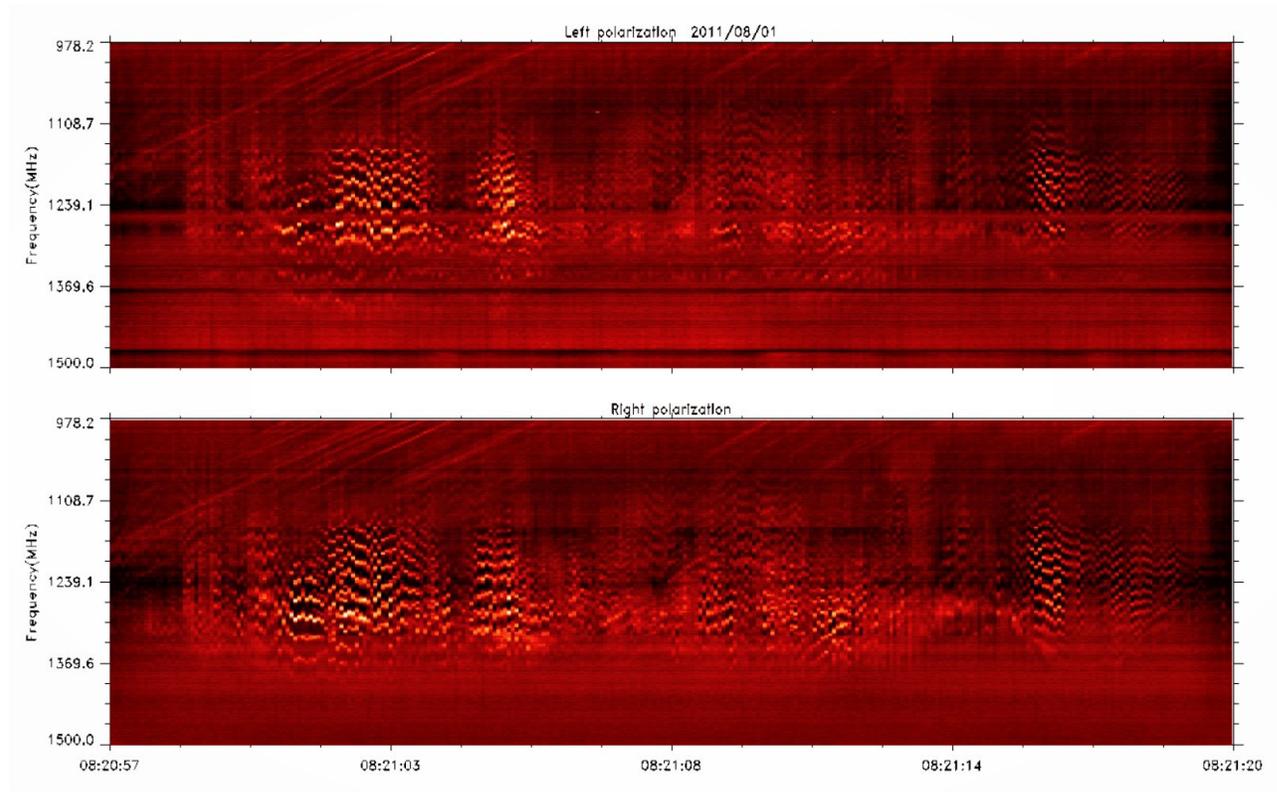

Fig. 2. ZP of pulsing regime in the decimeter wave range in the left and right polarization channels of the Yunnan observatory spectrograph (YNAO, Kunming, China). Fiber bursts with the direct frequency drift limit the ZP from the LF spectrum edge. The emission is weakly polarized with the right sign.

The location of the decimeter radio source of the zebra was identified from the appearance of new burst loops in the region close to the tail spot. However, the complex sporadic character of the ZP, as well as the appearance of the ZP in the microwave range (around 3000 MHz), remained beyond the authors' scope. In this work, we analyze the efficiency of alternative mechanisms and demonstrate the possible explanation of the main features of ZPs on the example of the same phenomenon on August 1, 2010, using a model with whistlers. It is impossible to consider the ten other ZP models in one publication; their importance requires additional analysis. A detailed description of the observational properties of the ZP and fiber bursts and their modern theoretical models are given in the literature (Chernov et al., 2015 and in two other reviews, Chernov, 2012 and 2016, which are freely accessible at http://www.izmiran.ru/~gchernov/).

## 2. OBSERVATIONS

The event on August 1, 2010, has already been the object of investigation in a few studies. Various aspects of the interaction between several coronal mass ejections (CMEs) in the interplanetary space were analyzed in four works (Temmer et al., 2012; Oliveiros et al., 2012; Schriver and Title, 2011; Liu et al., 2012). Liu et al. (2011) presented clear observations and analyzed the propagation of fast magnetosonic (FMS) waves from the region of a flare using films in the extreme ultra violet range based on the Solar Dynamic Observatory/Atmospheric Imaging Assembly (SDO AIA) 171 Å data. On the basis of these observations it was then shown (Karlický and Rybak, 2017) that FMS waves induced radio emission pulses in the range 25−2000 MHz with periods of ~181 s due to modulation of the acceleration processes of fast particles.

A flare with the importance of C3.2 started in AR 11092 (N13 E23) at 07:25 with a maximum at 08:57 and continued up to 10:05 UT. No clear type-II meter range bursts were observed (the IZMIRAN and Hiraiso spectra), although two CME occurred.

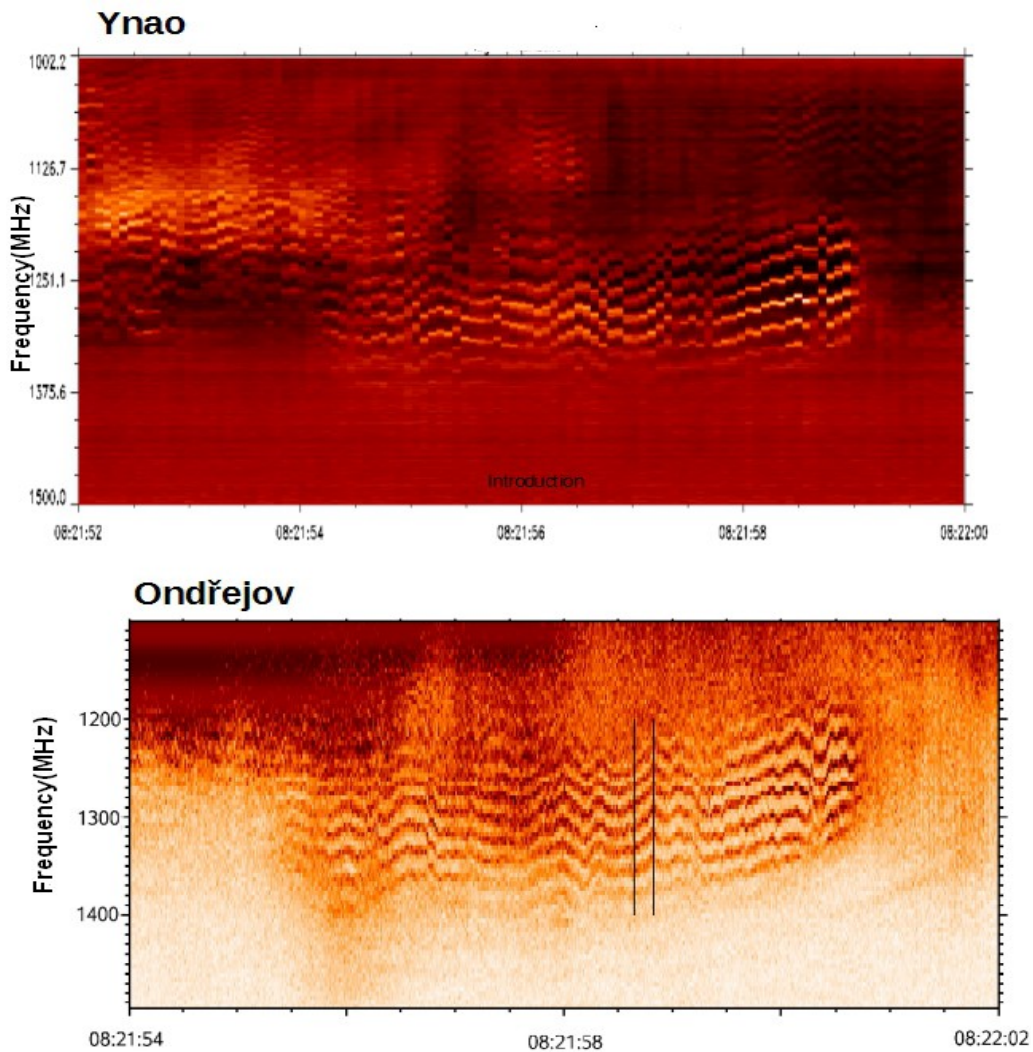

Fig. 3. Increased YNAO spectrum fragment of 8 s long as compared with the spectrum from the Ondřejov observatory (Yasnov et al., 2016) in the decimeter wave range. All zebra stripe details with wave-like frequency drift coincide from two observatories located at a distance of almost 8000 km, which confirms the solar nature of the ZP. The time resolution of the Ondřejov spectra is almost one order of magnitude higher, and the superfine structure of stripes with millisecond duration can be seen.

Two radio emission maxima were observed: after 0800 with left polarization and after 0900 UT with right polarization in the microwave range. Nearly unpolarized emissions were observed in the decimeter range. A type-II decameter burst was observed after the second maximum at 0908 UT. It is noteworthy that Oliveiros et al. (2012) consider this burst to be a result of the interaction between two CMEs.

Figure 1 shows a radio burst based on the data of the IZMIRAN spectrograph in the meter range (25−270 MHz). The burst resembles slowly drifting continuum. No clear stripes of the type-II burst are seen, which coincides with the Hiraiso spectrum. Pulsations with a period of ~3 min (Karlický and Rybak, 2017) are clearly seen on the upper general spectrum of the phenomenon. Pulsations with a period of ~5 s are clearly demonstrated in the spectrum of increased size in the range 45−90 MHz.

Figure 2 shows a zebra pattern (ZP) in the decimeter wave range in the pulsing regime in the right and left polarization channels of radio spectrograph from the YNAO observatory (Kunming, China). Fiber bursts with a direct frequency drift limit the ZP from the LF edge of the spectrum. The emission polarization is of the right sign and very weak. The pulsations of the ZPs are not strictly periodical and are almost instantaneous in the frequency band 1050−1370 MHz. The frequency separation between the stripes is not always the same in pulsations, but it notably increases with frequency from 15−20 MHZ to 35−40 MHz. The stripes become chaotic in the transition from one pulsation to another and reveal the splitting of one stripe into two. The frequency drift of the stripes is generally wave-shaped, but sometimes a sharp transition to the reverse (sawtoothed) type occurs. After 08:21:11 UT the stripes have an almost constant negative drift and resemble fiber bursts, especially at the HF edge of the spectrum.

Figure 3 shows a comparison of the YNAO spectra and the spectra from the Ondřejov observatory with a duration of 8 s approximately in the same frequency range. The data on zebra stripes with wave-shaped frequency drift coincide in details from both observatories, which are separated almost by 8000 km; this confirms the solar nature of the ZP. The time resolution of the Ondřejov observatory data (0.01 s) is approximately one order of magnitude higher than of the YNAO observatory.

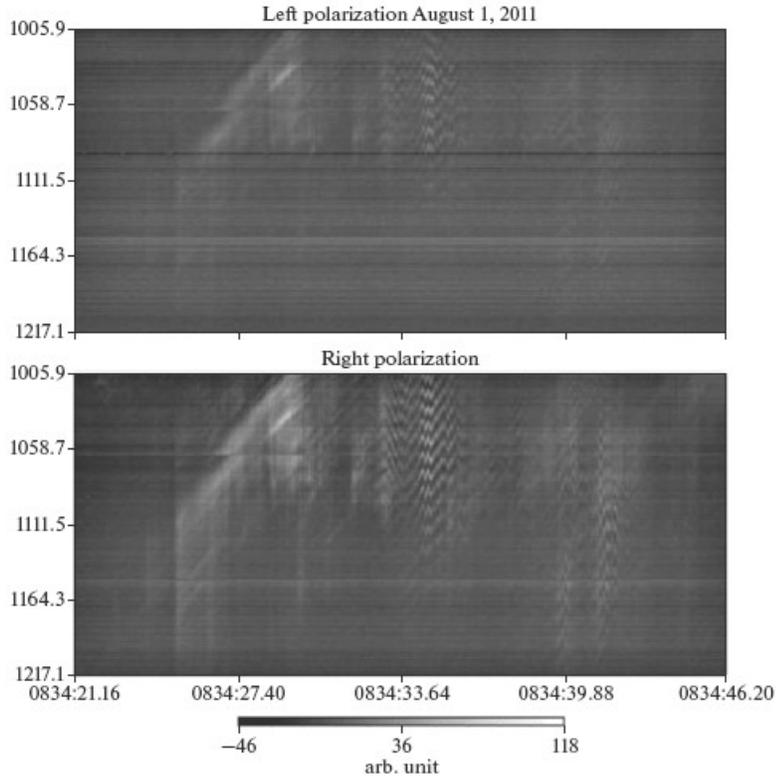

Fig. 4. Continuation of the ZP in the pulsing regime and fiber bursts based on the data of the YNAO observatory after 08:34 UT. The polarization of the right sign gradually increases and becomes strong.

The main time variations were manifested in the increase of the right polarization of the ZP and in the decreased frequency (Fig. 4), while the pulsing character and sawtoothed frequency drift remained. This is possibly related to the elevation of the radio source together, with the burst loop. By 08:39 UT a series of fiber bursts with strong right polarization appeared (Fig. 5). We note that the fiber bursts in the meter range at this moment were accompanied with pulsations (lower spectrum in Fig. 1), but the period of the existence of fiber bursts was one order of magnitude shorter (0.5 s).

For the completeness of analysis, we note that the ZP was also observed almost from the beginning of the event in the microwave range (Fig. 6). Unlike the decimeter range, we see here a restriction of the zebra by fiber bursts at the HF edge of the spectrum. In addition, no pulsing character of the stripes is observed. Their parameters continuously transform for 10 s; as the frequency increases, they practically transform to fiber bursts. The strong polarization of the left sign may be related to the location of this microwave source over the tail spot.

After the intensification of the burst at 09:08 UT, a new microwave source of the ZP appeared with strong right polarization (Fig. 7). Below, we will see that a burst intensification of brightness was observed over the leading spot during this time.

The further dynamics of the burst process is related to the emission of two families of fiber bursts with opposite frequency drift (Fig. 8) at the LF edge of the ZP spectrum.

According to the SDO/AIA 171 Å film data, the active elevation of flare loops and ejections with a predominant direction from the leading spot of the southern magnetic polarity were observed at the moments of the appearance of complex fine structure at 08:16−08:20 UT.

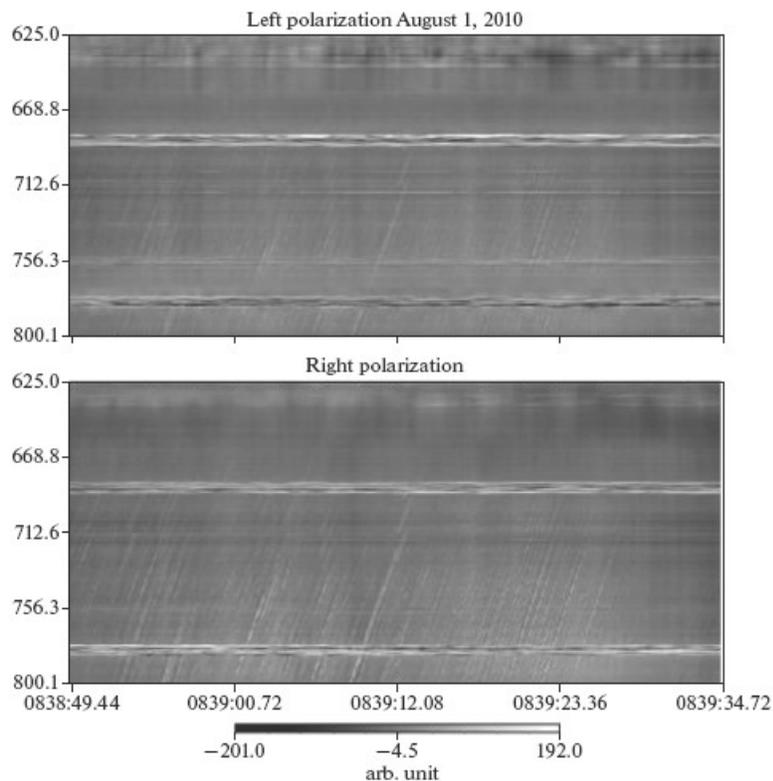

Fig. 5. Series of fiber bursts with direct frequency drift under strong right polarization (both spectra are based on the data of the YNAO observatory).

Therefore, we can suppose that it is most likely that the radio source was located at the top of the rising loops (denoted with thin dotted lines in Fig. 9), which led to very weak emission polarization of the right sign in the decimeter range. However, at 08:16, the polarization of the ZP in the microwave range was strong and had the left sign; hence, its source was located over the tail spot. An active flare loop was observed precisely there in the period 07:40−08:20 UT (it is shown with an arrow in the upper part of Fig. 9).

The degree of right-sign polarization gradually increased in the decimeter range. For example, at 08:34−08:39 UT a ZP and fiber bursts appeared at frequencies of 800−700 MHz already with strong right-sign polarization (Figs. 4 and 5). Analysis of the dynamics of flare loops in the SDO/AIA 171 Å film shows that the elevation of the flare loops over the leading spot was terminated by this moment. Even more, the upper loop started to descend, and by 08:39 UT it began to intersect with the lower loop. A bright flare node appeared at the place of their intersection, which is indicated with an arrow in Fig. 10. It is likely that we observed magnetic reconnection of the loops with the shear. After the second maximum of the burst, the fiber bursts and a sporadic ZP were also observed in the microwave range with strong right-sign polarization (Figs. 7 and 8). The two families of fiber bursts at the LF edge of the spectrum with the opposite frequency drift in Fig. 7 may be associated with the acceleration of the particles in the new magnetic island over the flare loop. The frequency interval between the initial frequencies of the fiber bursts (<2600 and 2800 MHz) may be determined by the size of the magnetic island and the acceleration of particles at two X-points in the upper and lower parts.

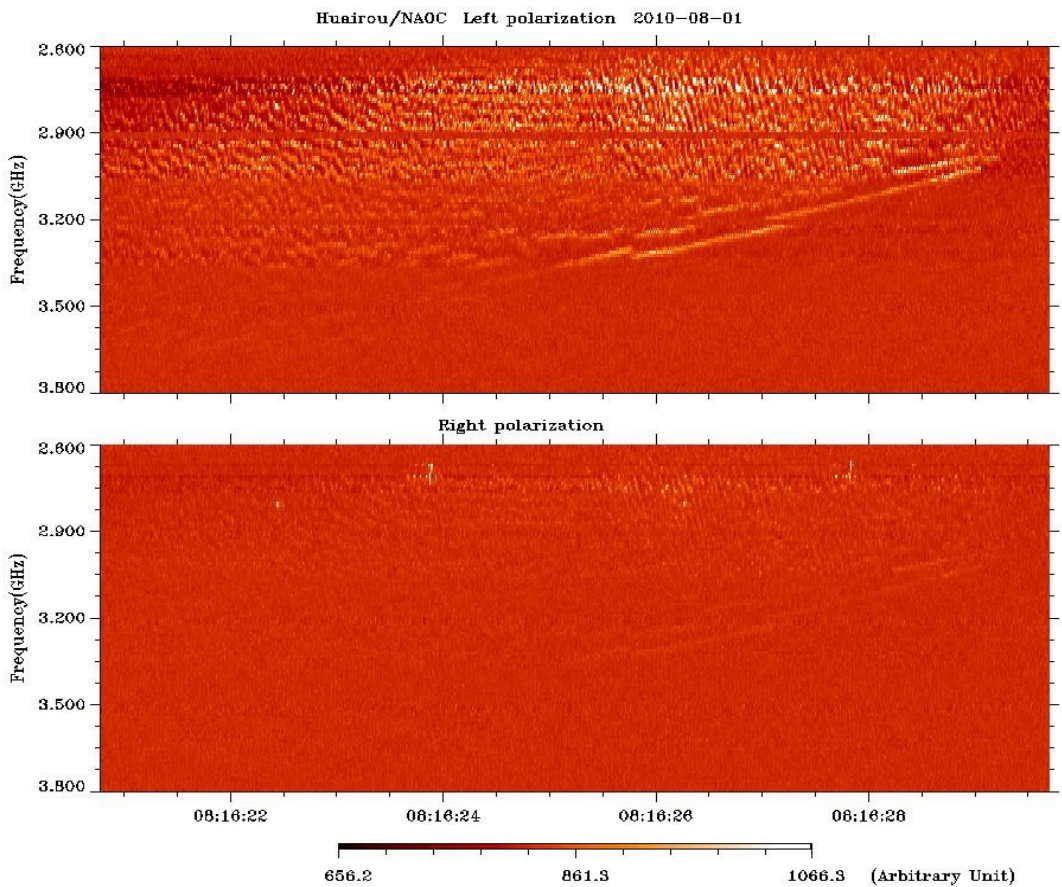

Fig. 6. ZP in the microwave range (based on the Huairou data, Beijing) in the beginning of the event with strong left polarization. Fiber bursts with frequency drifting from 3500 MHz to 3000 MHz limit it from the HF edge of the spectrum.

## 2.1. Conclusions Based on Observations

After the second maximum of radio burst at 09:08 UT, which accompanied the appearance of the second CME, helmet-shaped loops with a magnetic island over them formed over the active NOAA region 11092 (see Figs. 1 and 2 in (Schriver and Title, 2011)). The ongoing acceleration of particles from the zero points of the magnetic field under and over the magnetic island allows an understanding of the complex combinations of a pulsing ZP and fiber bursts.

It is likely that the magnetic island existed after the first CME over the node of magnetic reconnection in Fig. 10; the source of the ZP and fiber bursts in the decimeter range during the time interval 08:16−08:57 UT may be related to it. The fiber bursts in the LF spectrum range in Fig. 2 are related to the particle acceleration at the upper zero point of the island, while the particles trapped in the magnetic island form the loss-cone distribution of velocities responsible for the ZP. However, a number of the peculiarities of this event require explanation.

—No ZP existed in the meter range, but they were found only at frequencies greater than 1100 MHz with a gap of approximately 1400–2600 MHz, and they further appeared up to 3200 MHz where the stripes disintegrate into spikes, which are restricted to fiber bursts with negative drift.

—Fiber bursts sometimes superimpose on the ZP after 08:00 UT or limit it from the LF side in the decimeter range (or from the HF side in the microwave range).

—The ZP appears in the decimeter range in the pulsing regime in the form of chaotic instantaneous columns (almost without drift) with a chaotic duration from 0.1 s to 6 s. The frequency drift of the bands in the columns is not stationary but sawtoothed; zebra pulsations almost continuously and abruptly switch to the new pulsations; the zebra stripes on the HF side diffuse beginning from frequencies of ~1300 MHz;

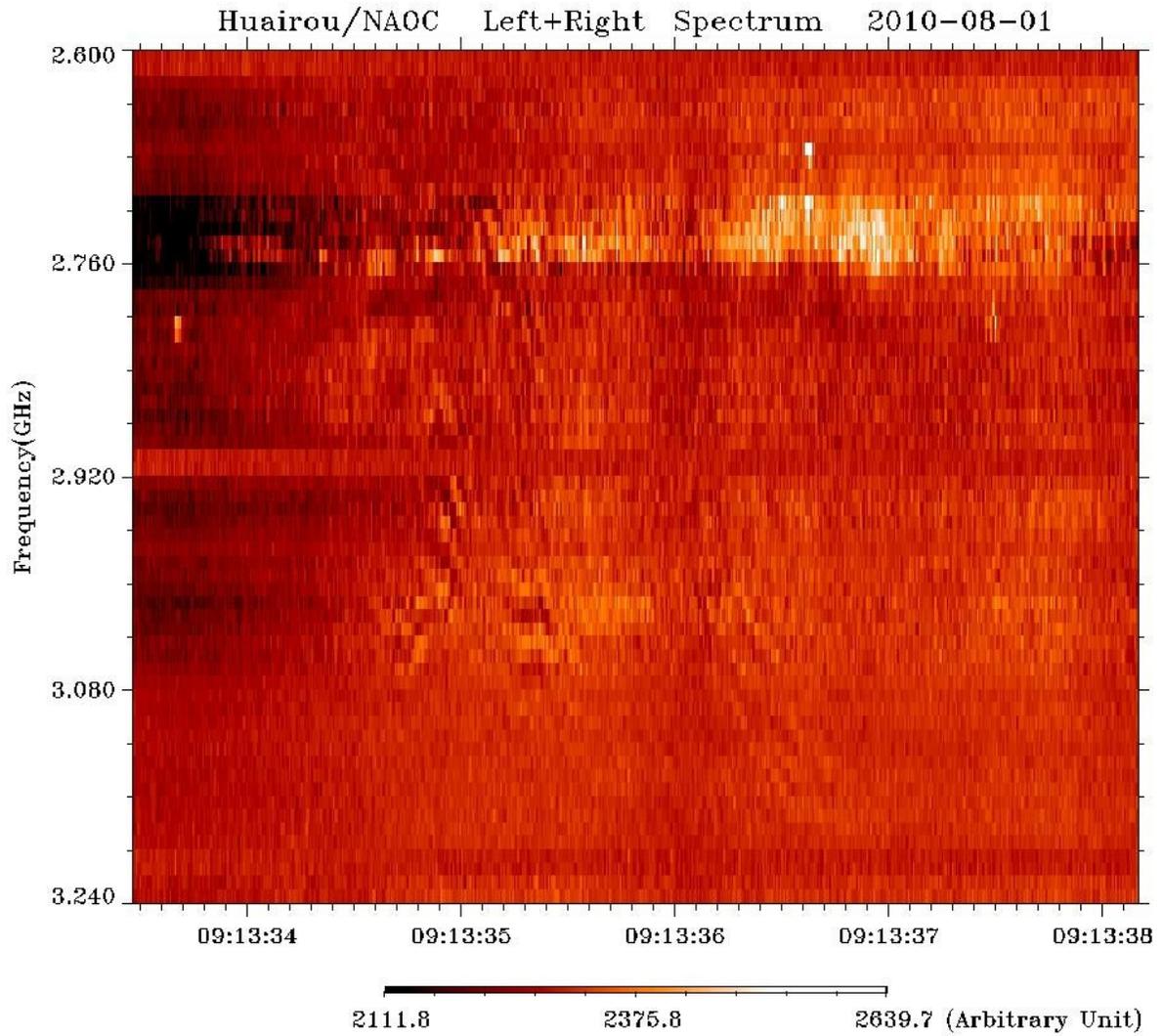

Fig. 7. Sporadic ZP at frequencies of 2900−3200 MHz (Huairou station, Beijing). Strong polarization of the right sign.

—The frequency separation increases with frequency (as usual); sometimes, one stripe splits into two.

—The ZP bands and continuum have a super fine structure.

## 3. DISCUSSION

ZPs in the decimeter range (Fig. 2) appear in a pulsing regime with different frequency drifts in individual pulsations (columns); however, we note that the stripes do not terminate and almost continuously (sometimes with attenuation) transform to the next column. The frequency drift changes each second or sometimes even faster; at 08:21:11 UT, the zebra stripes transform to fiber bursts with an approximate duration of four seconds.

In the ZP model, we have to suppose under DPR conditions that the magnetic field should sharply change several times in one second and that the ratio of the magnetic field gradients and density should change several times per second, which is hardly realized. The DPR levels cannot instantaneously appear and vanish in a large interval of altitudes. The uplift of a new magnetic flux and the formation of new magnetic loops are slow processes. Such variations at the altitudes of the decimeter range can be related only to the propagation of Alfven or fast magnetic acoustic waves from a flare. The ZP occupies the ~300 MHz spectral range, which can be determined by the size of the generated magnetic island; it may be as large as tens of thousands kilometers in any density model. Slow variations in the plasma parameters from the propagating waves are not able to change them almost instantaneously over such an interval of altitudes. A similar conclusion was made in the ZP analysis at frequencies of 7−7.5 GHz (Chernov et al., 2012).

Analysis of the spectra shown in Figs. 2 and 4 suggests that a pulsing process exists in the radio source that forms the ZP stripes. Such a process in the model with whistlers is related to the pulsing interaction of plasma Langmuir waves with whistlers. The change in the frequency drift of the stripes is related to the switching of whistler instability from the normal Doppler effect to the anomalous one due to the diffusion of fast particles on the whistlers. Figure 11 shows the possibility of such switching of whistler instability (Chernov, 1990; Chernov, 1996). A gradual switching of whistler instability from the dominant contribution of the normal Doppler effect to the anomalous one occurs according to the operator $\Lambda$ sign in the relation for the instability growth rate $\gamma$ (Gendrin, 1981; Bespalov and Trakhtengerts, 1980), which depends on the signs of the derivatives of distribution functions $F$ with respect to the longitudinal $V_{\parallel}$ and transversal velocities $V_{\perp}$:

$$\Lambda = (s\omega_{Be}/\omega V_\perp)(\partial/\partial V_\perp) + (k_{||}/\omega)(\partial/V_{||}) \quad |V_{||} = (\omega - s\omega_{Be})/k_{||}.$$

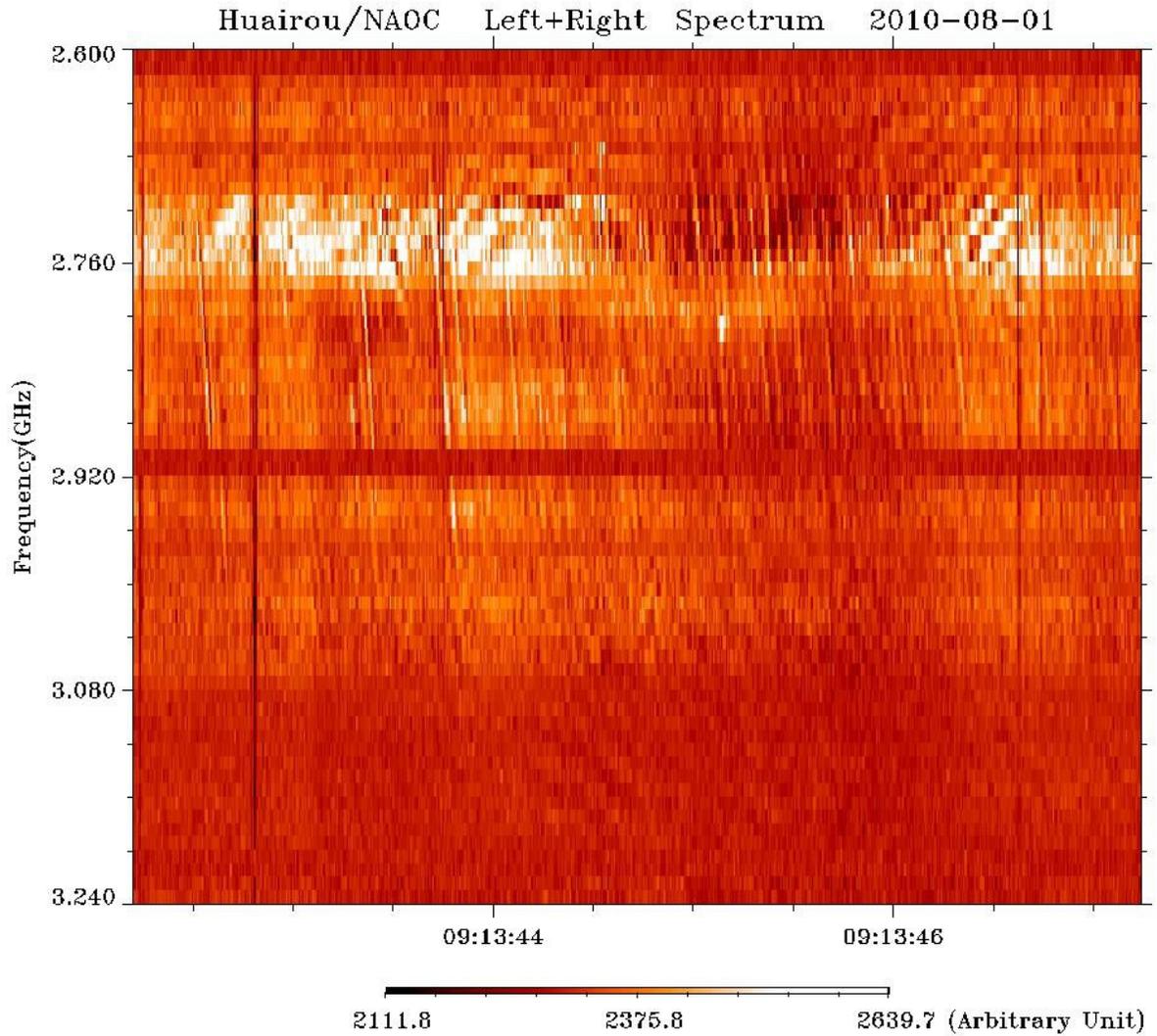

Fig. 8. Two series of fiber bursts of different scales with the direct and reverse drift and ZP at the HF edge of the spectrum at 3–3.2 GHz (Huairou station, Beijing). Strong polarization of the right sign. At 08:16 UT the polarization of the left sign was strong.

Operator $\Lambda$ has a clear physical sense of the derivative with respect to energy $E$ along the diffuse curve ($D$ in Fig. 11).

It is known that, under the conditions of the normal Doppler effect ($s = +1$ in the cyclotron resonance $\omega - k_{||}V_{||} - s\omega_{Be} = 0$), the particles and waves have opposite directions, while they have the same direction under the conditions of the anomalous Doppler effect ($s = -1$) but at a greater angle to the magnetic field. Resonance switching operates with a periodicity determined by the diffusion time, which is of the order of one second in the solar corona (Bespalov and Trakhtengerts, 1980). This effect slowly changes the group velocity of whistlers and thus changes the frequency drift of the ZP stripes. This effect was called bunch rotation due to diffusion (Shapiro and Shevchenko, 1987). This effect is also related to the fan instability in the evaluation of tokamak plasma (Parail and Pogutse, 1981).

The additional ejection of an electron beam almost instantaneously switches the whistler instability to the anomalous resonance. The intermediate position of the bunch in Fig. 11 forms the conditions for splitting of the zebra stripe into two, simultaneously in both resonances. This splitting effect can be seen, for example, in Fig. 2 at 08:21:01 UT.

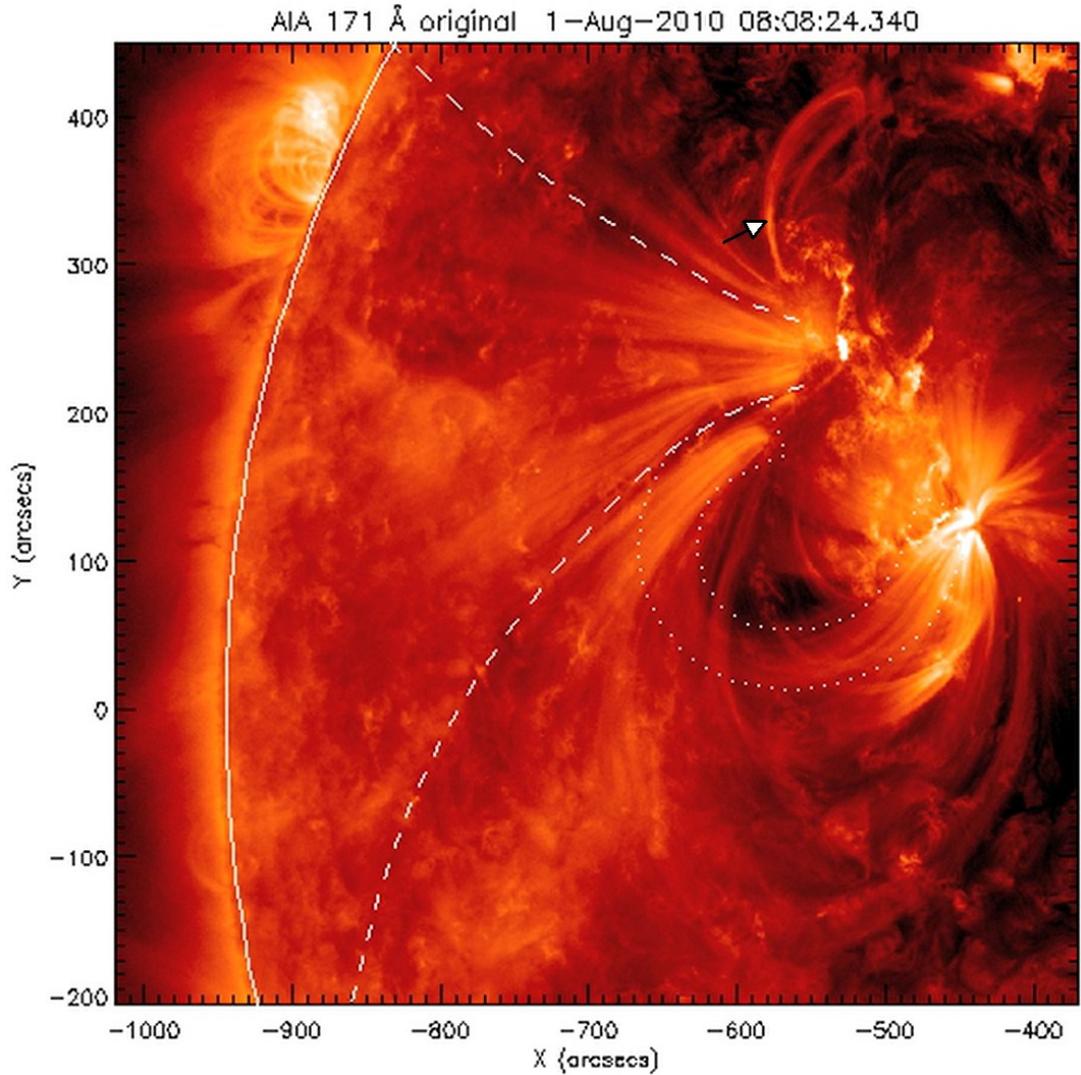

Fig. 9. Image of the active region in the ultra violet range based on the KA SDO/AIA 171 Å data (fragment of Fig. 1 from (Oliveiros et al., 2012)). The decimeter radio source may have been located at the tops of rising loops denoted with thin dotted line. The arrow in the upper part denotes the active flare loop, which can be associated with the radio source of the ZP in the microwave range at 08:16 UT.

Fiber bursts are usually formed during whistler instability under ordinary Doppler effect when the transversal velocities dominate in the beam, while ZPs dominate during anomalous resonance. Therefore, the transition of ZPs to fiber bursts (and reverse) in Fig. 2 at 08:21:11 and 08:21:17 UT (and many times at the other time moments) is related to such switching. The switching of whistler instability with the rotation of the group velocity of whistlers should be accompanied by a synchronous change in the direction of the frequency drift of ZP stripes on the spectra with a change in the direction of the spatial drift of their radio sources. Such a synchronicity has been repeatedly observed in the meter range (Chernov, 2005; Chernov, 2006).

The distribution function gradually becomes flatter due to the diffusion on whistlers, which leads to decay of the instability of the plasma waves in the volume of the wave packet of whistlers. This decay is the main cause of the formation of dark stripes at the LF edge of fiber bursts and ZPs.

The superfine structure of zebra stripes in the model with whistlers (Figs. 3, 7, 8) can be naturally explained within the pulsing regime of the interaction between whistlers and ion-sound waves (Chernov et al., 2003).

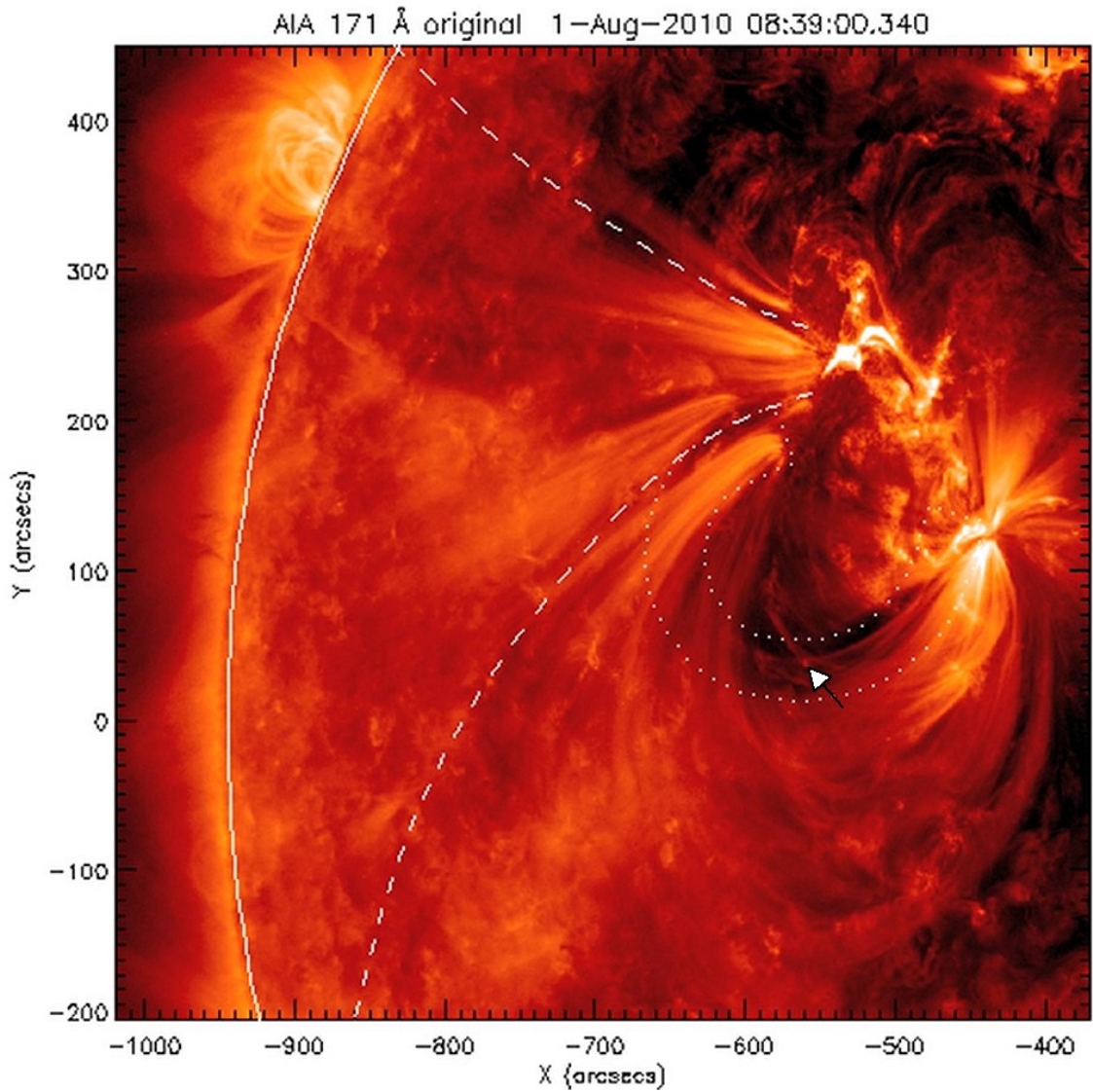

Fig. 10. Image of the active region in the ultra violet range based on the SDO/AIA 171 Å data (fragment of Fig. 1 from Oliveiros et al, 2012). The moment of magnetic reconnection of the rising loops with increased brightness of the location of their crossing is denoted by the arrow. This is the moment of the series of fiber bursts in the decimeter range with strong right polarization shown in Fig. 5.

.

Narrow-band chains (ropes) of fiber bursts are related to the periodical instability of whistlers between the fronts of fast shock waves in the region of magnetic reconnection (Chernov, 1990).

4. CONCLUSIONS

Due to the joint observations using Chinese spectrographs at the YNAO and Huairou observatories and at the Institute of Terrestrial Magnetism, Ionosphere, and Radio wave Propagation, Russian Academy of Sciences, we demonstrated a wide variety of sporadic ZPs in the decimeter and microwave ranges during the event on August 1, 2010 (Figs. 2–8). The ZPs were observed in a pulsing regime with a sharp (sawtoothed shape) variation in the direction of the drift of stripes. Various combinations of zebra stripes with fiber bursts and their continuous transition from one form to another were observed. The zebra stripes reveal a superfine structure of millisecond duration.

The general properties of the fine structure of radio emission are usually related to the dynamics of the burst process: the sign and degree of the circular polarization, various combinations of the ZPs and fiber bursts, and the pulsing character on the dynamical spectrum.

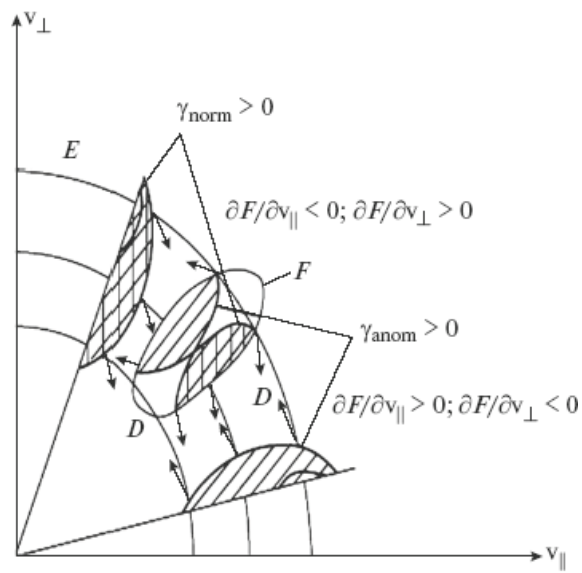

Fig. 11. Switching of whistler instability from the normal Doppler effect to the anomalous one according to Gendrin (1981) (fragment of Fig. 2 from Chernov (1990)). The instability region with the anomalous Doppler effect is denoted by ordinary shading, and the one with the normal resonance is denoted by double shading; D is the direction of diffuse curves.

The main details of the sporadic ZP in the event on August 1, 2010, were explained within the model of the ZPs and fiber bursts during the interaction of the plasma waves with whistlers. The main variations in the ZPs stripes are caused by the scattering of fast particles on whistlers, which leads to the switching of whistler instability from the normal Doppler effect to the anomalous one.

ACKNOWLEDGMENTS

The authors thank their Chinese colleagues Chenming Tan and Guannan Gao from the National Astronomy Observatory of China (NAOC). This work was supported by the Russian Foundation for Basic Research (project nos. 17-02-00308, 17-52-80064). We thank the teams of the YNAO, Ondřejov, and SDO/AIA observatories, who allowed open access to their data. This work was supported by the Program of the Presidium of the Russian Academy of Sciences, grant no. 7.